\documentclass[preprint,12pt]{elsarticle}

\usepackage{graphicx}

\journal{New Astronomy}

\begin{document}

\begin{frontmatter}



\title{Time Delay in Robertson-McVittie Spacetime
and its Application to Increase of Astronomical Unit}


\author{Hideyoshi ARAKIDA}

\address{School of Education, Waseda University\\
1-6-1, Nishi-Waseda, Shinjuku-Ku, Tokyo 169-8050, Japan\\
E-Mail : {\tt arakida@edu.waseda.ac.jp}}

\begin{abstract}
We investigated the light propagation by means of the Robertson-McVittie 
solution which is considered to be the spacetime around the gravitating
body embedded in the FLRW 
(Friedmann-Lema{\^i}tre-Robertson-Walker) background metric.
We concentrated on the time delay and derived the correction terms
with respect to the Shapiro's formula. To relate with the 
actual observation and its reduction process, we also took account of 
the time transformations; coordinate time to proper one, 
and conversely, proper time to coordinate one. 
We applied these results to the problem of increase of astronomical 
unit reported by Krasinsky and Brumberg (2004). 
However, we found the influence of the cosmological expansion 
on the light propagation does not give an explanation of observed 
value, $d{\rm AU}/dt = 15 \pm 4$ [m/century] 
in the framework of Robertson-McVittie metric.
\end{abstract}


\end{frontmatter}

\section{Introduction\label{intro}}
The appearance of radar/laser gauging techniques 
and spacecraft ranging has enabled us 
to measure the distance from the Earth to other inner planets 
and the moon directly and precisely. Recent years, these techniques 
have been drastically improved, 
for instance, the planetary radar measurement achieves the 
observational accuracy of the interplanetary distance within a few 
100 [m], the spacecraft ranging a few [m] and the lunar laser ranging 
a few [cm]. In these observations, the round-trip time of 
light/signal is measured by the atomic clocks on the Earth. 
Nowadays, the advent of laser cooled atomic clocks has led to the
improvement of realization of SI second and it is expected that
this advance will go on. Therefore the arrival time measurement is 
the most accurate observation
among the all kinds of astronomical and astrophysical ones. 

Due to the improvement of these techniques, it is important to 
develop the rigorous light propagation model which contains the 
various general relativistic and relating effects, and this subject 
is actively investigated by many authors, e.g. 
\cite{shapiro1964,richter1983,hellings1986,kopeikin1997,kopeikin1999,
kopeikin2002,klioner2003a,klioner2003b,defellice2004,defelice2006,asada}
and references therein.
These theoretical developments also play a crucial role in testing 
the gravitational theories \cite{will1,will2}. 

On the other hand, an arrival time measurements have also 
devoted to the improvement of lunar and planetary ephemerides,
such as DE \citep{standish2003}, EPM \citep{pitjeva2005}, 
VSOP \citep{bretagnon1988}, and INPOP \citep{fienga2008}, and the
determination of astronomical constants. Especially among these
constants, the astronomical unit (of length) AU is one of the 
fundamental and important one which gives the relation of 
length units; [AU] of astronomical system of units and [m] of SI ones. 
AU is currently determined from the arrival time measurement data of
planetary radar and spacecraft ranging within the 
accuracy of 0.1 [m] or 12 digits 
level as $1 ~[{\rm AU}] = 1.495978706960 \times 10^{11} \pm 0.1 
~[{\rm m}]$ \cite{pitjeva2005}. Therefore from the point of view of 
the fundamental astronomy, the development of a
accurate light propagation model is of great importance.

Until now, the light propagation formulae have been derived in terms 
of the post-Newtonian approximation,
$g_{\mu\nu} = \eta_{\mu\nu} + h_{\mu\nu}, ~|h_{\mu\nu}| \ll 1$ in which 
the background metric $\eta_{\mu\nu}$ is the static Minkowski one. 
However, it is interesting to examine the influence of cosmological
expansion on the light/signal propagation  since the accuracy of 
astronomical observation is rapidly growing. 
Of course, it is presently hard to observe such a influence  
in the solar system experiments.
Nevertheless, the progress of these or relating measurement techniques 
may enable to detect its trace in the future date.

In this paper, we will study the light propagation passing near the 
massive star as the Sun which is embedded in the cosmological 
expanding background.
We will adopt the Robertson-McVittie spacetime introduced in section 
\ref{rm-metric}, and concentrate on the gravitational 
time delay and derive the extra delay due to the 
effect of cosmological expansion with respect to the 
standard formula by Shapiro in section \ref{time-delay}. 
Further in order to relate with the actual observation and its 
reduction process, we will take account of 
the time transformations in section \ref{tt}; 
from coordinate time to proper one in subsection \ref{ctp}, 
and conversely, from proper time to coordinate one in subsection 
\ref{ptc}. As the application of results, we will consider the problem of 
increase of astronomical unit recently reported by \cite{kb2004} 
in section \ref{au}.
\section{Robertson-McVittie Spacetime\label{rm-metric}}
The standard form of Robertson-McVittie metric is given by
\citep{robertson,mcvittie,jarnefelt1,jarnefelt2,jarnefelt3,
carrera,nolan1,nolan2},
\begin{eqnarray}
 ds^2 &=& -\left(
	  \frac{1 - \frac{GM}{2c^2ra(t)}}{1 + \frac{GM}{2c^2ra(t)}}
	 \right)^2 c^2dt^2 
\nonumber\\
 & &+ \left(1 + \frac{GM}{2c^2ra(t)}\right)^4
 a^2(t)(dr^2 + r^2d \theta^2 + r^2\sin^2 \theta d\phi^2),
 \label{mcvittie}
\end{eqnarray}
where $G$ is the gravitational constant, $M$ is the mass of central
body, $c$ is the speed of light in vacuum, and $a(t)$ is a scale factor. 
(\ref{mcvittie}) is regarded as the spacetime around the point 
mass singularity embedded in the FLRW background metric, and 
coincides thoroughly with the Schwarzschild solution in the 
isotropic coordinate when $a(t) = 1$, and the FLRW model for 
curvature parameter $k=0$ when $M = 0$. 

However, recent dynamical model of planetary motion (EIH 
equation of motion), the various light propagation models
cited in the previous section and other observational models 
in the solar system are formulated not in the comoving coordinate system as 
(\ref{mcvittie}) but in the barycentric celestial reference system 
(BCRS) or corresponding to BCRS based on the post-Newtonian 
framework \cite{soffel}. And the various astronomical constants are
also derived in BCRS or corresponding to BCRS; for example 
the secular increase of AU,
$d{\rm AU}/dt$ that we discuss in section \ref{au}
is evaluated based on barycentric dynamical time (TDB)
\cite{krasinsky}.
Therefore in order to discuss these effects and cosmological one in 
the same framework as far as possible, 
it seems to be more adequate to take the 
reference system as close as possible to BCRS. 
Hence we adopt the following radial transformation and
convert (\ref{mcvittie}) into the nearly proper coordinate system
\cite{robertson,jarnefelt1,jarnefelt2,jarnefelt3,carrera,nolan1,nolan2},
\begin{equation}
 R = a(t)r\left(1 + \frac{GM}{2c^2ra(t)}\right)^2.
  \label{radial-transform}
\end{equation}
We notice that the matching of BCRS and cosmological reference system
is important problem in the fundamental astronomy and this issue is 
discussed by \cite{ks2004,kopeikin2007}.
Using (\ref{radial-transform}), (\ref{mcvittie}) is rewritten as,
\begin{eqnarray}
 ds^2 &=& -\left(1 - \frac{2GM}{c^2R}\right)c^2dt^2
  \nonumber\\
  & &+
  \left(\frac{dR}{\sqrt{1 - \frac{2GM}{c^2R}}} - \frac{H(t)}{c}Rcdt\right)^2
  + R^2(d\theta^2 + \sin^2\theta d\phi^2),
\label{mcvittie-2}
\end{eqnarray}
here $H(t) = \dot{a}(t)/a(t)$ is a Hubble parameter and the 
overdot denotes the time derivative.
If $H(t)$ does not change, namely, the Hubble constant at present, 
$H(t) \rightarrow H_0 = (h_0/3.08) \times 10^{-17}$ [1/s], 
$h_0 = 0.7$, 
then we can introduce the time transformation to remove 
$dtdR$ term in (\ref{mcvittie-2}) \cite{robertson,jarnefelt1},
\begin{equation}
 cT = ct + \frac{H_0}{c}
  \int\frac{RdR}{\left(1 - \frac{2GM}{c^2 R} -
		  \frac{H_0^2}{c^2}R^2\right)
  \sqrt{1 - \frac{2GM}{c^2 R}}}.
  \label{time-transform}
\end{equation}
In a practical sense, the Hubble parameter $H(t)$ varies with time. 
Nonetheless, when focusing on the solar system experiments, the actual 
round-trip time of light/signal and the time interval in which the 
observational data is stored are much shorter, at most $t \sim 100$
[yr], than the age of Universe, $T_{\rm U} \sim 10^7$ [yr]. 
Thus, let us assume $H(t)$ changes adiabatically as,
\begin{equation}
 H(t) \simeq H_0 + \left.\frac{dH}{dt}\right|_0 (t - t_0).
\end{equation}
It is suited to put,
$\left.dH/dt\right|_0 \sim H_0/T_{\rm U} \sim 10^{-24}$ 
~[1/(s $\cdot$ yr)], hence, it follows,
\begin{equation}
 H_0 \sim 10^{-17} > \left.\frac{dH}{dt}\right|_0 \times 
  100 ~{\rm [yr]} \sim 10^{-22} ~~[1/{\rm s}].
  \label{hubble_0}
\end{equation}
From (\ref{hubble_0}), $H_0$ produces the dominant effect of
cosmological expansion. 
Substituting (\ref{time-transform}) into (\ref{mcvittie-2}) 
and limiting the equatorial motion $\theta = \pi/2$, we obtain,
\begin{eqnarray}
 ds^2 = -\left(1 - \frac{2GM}{c^2r} - \frac{H_0^2}{c^2}r^2\right)
  c^2dt^2 
  + \left(1 + \frac{2GM}{c^2r} + \frac{H_0^2}{c^2}r^2\right)dr^2
  + r^2 d\phi^2,
  \label{base-metric}
\end{eqnarray}
where we expanded the coefficient of $dr^2$ and 
replaced $T \rightarrow t$ and $R \rightarrow r$, respectively.

Here we note that the assumption 
$H(t) = \mbox{const.}$ leads Robertson-McVittie solution to 
Schwarzschild-de Sitter model, see e.g. \cite{ks2004,robertson}.
Friedmann equation is,
\begin{eqnarray}
 \left(\frac{\dot{a}}{a}\right)^2
 + \frac{kc^2}{a^2} = \frac{8\pi G}{3}\rho + \frac{\Lambda}{3}c^2
\end{eqnarray}
where $k$ is the curvature parameter, $\rho$ is the density of 
universe, and $\Lambda$ is the cosmological constant.
When the cosmological background is the de-Sitter universe, 
$\rho = 0$ and $k=0$, we have,
\begin{eqnarray}
  H(t) = \frac{\dot{a}}{a} = c\sqrt{\frac{\Lambda}{3}} =
   \mbox{constant},
   \label{de-sitter}
\end{eqnarray}
in this case $a(t) = \exp (c \sqrt{\Lambda/3} t)$.
Replacing $H_0$ by (\ref{de-sitter})
and inserting (\ref{radial-transform}) and (\ref{time-transform}),  
into (\ref{mcvittie}), we can recover the Schwarzschild-de Sitter 
solution.

As a consequence, although our metric (\ref{base-metric}) is equivalent
to Schwarzschild-de Sitter model, we will begin with this metric
as the first attempt of our investigation. 
\section{Time Delay in Robertson-McVittie Spacetime\label{time-delay}}
The world line of light/signal is the null geodesic, 
$ds^2 = 0$, then from (\ref{base-metric}) it results,
\begin{eqnarray}
 ct(r, r_0) 
  &=& \int_{r_0}^rdr^{\prime}
  \sqrt{\frac{1 + \frac{2GM}{c^2 r^{\prime}} 
  + \frac{H_0^2}{c^2}r^{\prime 2}}
  {1 - \frac{2GM}{c^2 r^{\prime}} - \frac{H_0^2}{c^2}r^{\prime 2}}
  }
  \nonumber\\
 & &\quad \times
  \left[
   1 - 
   \frac{1 - \frac{2GM}{c^2 r^{\prime}} 
   - \frac{H_0^2}{c^2}r^{\prime 2}}
   {1 - \frac{2GM}{c^2 r_0} - \frac{H_0^2}{c^2}r_0^2}
   \left(\frac{r_0}{r^{\prime}}\right)^2
  \right]^{-1/2},
\end{eqnarray}
in which $r_0$ is the closest point (impact parameter) between the 
light/signal and the central body. Remaining 
${\cal O}(GM/c^2, H^2_0/c^2)$ terms in the integrand, we obtain,
\begin{eqnarray}
 t(r, r_0) &=& t_{\rm 1PN}(r, r_0) + t_{\rm Cosmo}(r, r_0),\\
 t_{\rm 1PN}(r,r_0)&=&
  \frac{1}{c}\sqrt{r^2 - r^2_0}
  \nonumber\\
 & &
  + \frac{GM}{c^3}
  \left[
   2\ln
  \left(\frac{r + \sqrt{r^2 - r^2_0}}{r_0}\right)
  +
  \sqrt{\frac{r - r_0}{r + r_0}}
  \right],
  \\
 t_{\rm Cosmo}(r, r_0) &=& \frac{H_0^2}{6c^3}\sqrt{r^2 - r_0^2}
  \left(2r^2 - r_0^2 \right),
\end{eqnarray}
here $t_{\rm 1PN}$ is the Shapiro time delay in the 1st 
post-Newtonian approximation, and $t_{\rm Cosmo}$ is the extra one 
caused by the cosmological expansion. If we assume the Earth $(E)$ 
and the objective planet/spacecraft $(R)$ are almost at rest during 
the round-trip of light/signal, then the net of round-trip time 
${\cal T}$ becomes,
\begin{equation}
 {\cal T} = {\cal T}_{\rm 1PN} + {\cal T}_{\rm Cosmo} = 
  2[t(r_E, r_0) + t(r_R, r_0)].
  \label{delay1}
\end{equation}
The time delay produced by the cosmological expansion is,
\begin{eqnarray}
 {\cal T}_{\rm Cosmo} =
  \frac{H_0^2}{3c^3}
  \left[
   \sqrt{r_E^2 - r^2_0}\left(2r_E^2 - r_0^2\right) 
   + \sqrt{r_R^2 - r^2_0}\left(2r_R^2 - r_0^2\right) 
  \right].
  \label{delay2}
\end{eqnarray}
\section{Time Transformations\label{tt}}
\subsection{Transformation from Coordinate Time to Proper Time\label{ctp}}
(\ref{delay1}) and (\ref{delay2}) give the time delay in the 
coordinate time. However, the measurement of round-trip time is 
actually carried out by the atomic clocks on the Earth 
which ticks the proper time $\tau$. Therefore, we must 
convert (\ref{delay1}) and (\ref{delay2}) into the observer frame.

To this end, it is sufficient to consider the equation of proper time
in the quasi-Newtonian approximation, 
\begin{equation}
 \frac{d\tau}{dt}
  =
  1 - \frac{v^2}{2c^2} - \frac{GM}{c^2 r} - \frac{H_0^2}{2c^2}r^2.
\label{propertime}
\end{equation}
Taking the orbital radius and the velocity of the Earth, 
$r_E$ and $v_E$, respectively, the measured round-trip time 
$\bar{\cal T}$ becomes, 
\begin{equation}
\bar{\cal T} = \left. \frac{d\tau}{dt} \right|_E {\cal T} =
\bar{\cal T}_{\rm 1PN} + \bar{\cal T}_{\rm Cosmo}.
\end{equation}
Making use of (\ref{delay1}), the part due to the cosmological 
expansion relating with $H_0$ is,
\begin{eqnarray}
 \bar{\cal T}_{\rm Cosmo}
  &=&
  \frac{H_0^2}{c^3}
 \left\{
   \frac{1}{3}
   \left[
    \sqrt{r_E^2 - r_0^2}\left(2r_E^2 - r_0^2\right)
    + \sqrt{r_R^2 - r_0^2}\left(2r_R^2 - r_0^2\right)
   \right]\right.
 \nonumber\\
 & &\qquad
  -
 \left.
 r^2_E
 \left(
 \sqrt{r_E^2 - r_0^2} +\sqrt{r_R^2 - r_0^2}\right)
  \right\}
 \nonumber\\
 & &-
  \frac{GMH_0^2}{c^5}r^2_E
  \left(
   2\ln \frac{r_E + \sqrt{r^2_E - r^2_0}}{r_0}
   +
   2\ln \frac{r_R + \sqrt{r^2_R - r^2_0}}{r_0}
   \right.\nonumber\\
 & &\qquad
   + \left.
   \sqrt{\frac{r_E - r_0}{r_E + r_0}}
   +
   \sqrt{\frac{r_R - r_0}{r_R + r_0}}
  \right)\nonumber\\
 & &-
  \frac{1}{3c^5}
  \left(\frac{v^2_E}{2} + \frac{GM}{r_E} + \frac{H^2_0}{2}r^2_E\right)
  H^2_0\nonumber\\
 & &\qquad \times
    \left[
    \sqrt{r_E^2 - r_0^2}\left(2r_E^2 - r_0^2\right)
    + 
    \sqrt{r_R^2 - r_0^2}\left(2r_R^2 - r_0^2\right)
   \right].
\label{dtau}
\end{eqnarray}
This is the extra time delay observed by the atomic clocks on 
the Earth. ${\cal O}(c^{-3})$ terms in 
first and second line are the leading ones due to the cosmological
expansion and ${\cal O}(c^{-5})$ ones from third to sixth line
are the coupling terms i.e. cosmological expansion and 
Shapiro delay, and so on. Hence the coupling terms are considered to be
next order of dominant terms in terms of post-Newtonian 
approximation. 
\subsection{Transformation from Proper Time to Coordinate Time\label{ptc}}
Next, we consider the transformation from proper time to 
coordinate time.
This transformation is also of importance when obtaining the 
position and velocity of planet/spacecraft (they are the reflectors of 
light/signal) referring the proper time $\tau$ of observer on the
Earth (practically, coordinated universal time, UTC or international 
atomic time, TAI). This is due to the fact in the ephemeris the 
position and velocity of celestial body 
is calculated as the function of the coordinate time $t$ 
which is the independent variable of the equation of motion.

Applying the Keplerian energy integral,
\begin{equation}
 \frac{1}{2}v^2 - \frac{GM}{r} = -\frac{GM}{2a_0},
\end{equation}
where $a_0$ is the semi-major axis of planetary orbit, and 
(\ref{propertime}) is rewritten as, assuming the Earth (observer)
moves along the circular orbit, $r = r_E = a_0$, 
\begin{equation}
 \frac{d\tau}{dt}
  =
  1 - \frac{3GM}{2c^2 r_E} - \frac{H_0^2}{2c^2}r_E^2.
\label{propertime2}
\end{equation}
By integrating (\ref{propertime2}) we have, 
\begin{eqnarray}
 \tau &=& \tau_{\rm 1PN} + \tau_{\rm Cosmo},\\
 \tau_{\rm 1PN} &=&
  \left(1 - \frac{3GM}{2c^2 r_E}\right)(t - t_0),
  \label{tauh1}\\
 \tau_{\rm Cosmo} &=& 
  - \frac{H_0^2r_E^2}{2c^2}(t - t_0),
  \label{tauh2}
\end{eqnarray}
here $\tau_{\rm 1PN}$ represents the relation between the 
coordinate time $t$ and the proper time $\tau$, 
sometimes called the time ephemeris 
(the accurate analytical treatment of $\tau_{\rm 1PN}$ is given by, 
e.g. \cite{moyer1981a,moyer1981b,fb1990}).
$\tau_{\rm Cosmo}$ is caused due to the cosmological expansion.
\section{Application to Increase of Astronomical Unit\label{au}}
In this section, let us apply above results to the increase of 
astronomical unit recently reported by Krasinsky and Brumberg 
(2004) \cite{kb2004}.
They found that from the analysis of high precision planetary radar 
and martian spacecraft ranging data (the arrival time measurement), 
the astronomical unit of length AU increases with respect to 
meters as $d{\rm AU}/dt = 15 \pm 4$ [m/century]. 
This reported values is about 100 times larger than 
the present determination error of AU, see \cite{pitjeva2005} and 
section \ref{intro} in this paper. 
The similar variation of AU is also corroborated by \citep{standish2005}.

Before entering the AU issue, we briefly summarize how the AU is 
derived from the analysis of observational data.
These days, the AU is determined based mainly on the planetary 
radar and the spacecraft ranging data since these data give the 
interplanetary distance $\ell_{\rm obs}$ directly and precisely as, 
$\ell_{\rm obs} = c t_{\rm obs} ~[{\rm m}]$ where 
$t_{\rm obs} ~[{\rm s}]$ is the observed round-trip time of 
light/signal. While, the planetary ephemerides provide the 
interplanetary distance $r_{\rm theo}$ in the unit of [AU]
as the theoretical value. Then
to compare $r_{\rm theo}$ in [AU] with observed $t_{\rm obs}$ in [s], 
$r_{\rm theo}$ is converted into $t_{\rm theo}$ in [s] as, 
\begin{equation}
 t_{\rm theo} \equiv r_{\rm theo}\frac{\rm AU}{c} = r_{\rm theo}\tau_A
  ~[{\rm s}]
\end{equation}
in which $\tau_A \equiv {\rm AU}/c$ is called the light time. 
Practically this step looks around the various effects, e.g. the 
relativity, the solar corona, and the troposphere, see 5.32 of
\cite{seidelmann2005}. 
Therefore the light propagation model plays a key role in deriving the 
AU. We also note that $r_{\rm theo}$ 
depends on a constellation of astronomical constants and
parameters via the equation of motion and some other relations. Then
AU is optimized and derived with other parameters simultaneously 
by the least square method, as satisfying $t_{\rm obs} = t_{\rm theo}$.

However, when Krasinsky and Brumberg rewrote $t_{\rm theo}$ as, 
\begin{equation}
t_{\rm theo} = \frac{r_{\rm theo}}{c}
 \left(
  {\rm AU} + \frac{d {\rm AU}}{dt}T
 \right),
\end{equation}
and fit to observational data, they discovered $d{\rm AU}/dt$ is 
non-zero and positive value $d{\rm AU}/dt = 15$ [m/century],
where $T$ is the time interval counted off from some initial epoch. 
Notice that estimated $d{\rm AU}/dt$ does not mean the expansion of 
planetary orbit and/or the increasing of orbital period of planet. 
According to Krasinsky \cite{krasinsky}, the observations do not 
show any traces of such kinds.
Further the determination error of inner planetary orbits in the latest 
planetary ephemerides is also smaller than the observed 
$d{\rm AU}/dt = 15$ [m/century], see e.g. Table 4 of \cite{pitjeva2005}. 
Therefore the observed $d{\rm AU}/dt$ may relate with
not the dynamical aspect of planetary motion but 
the light/signal propagation.

Therefore, let us examine if the extra time delay derived 
in previous sections can explain the observed $d{\rm AU}/dt$.
We first apply (\ref{dtau}) and evaluate the Earth-Mars ranging 
concretely. We suppose the Earth ($E$) and the Mars ($R$) move along 
the circular orbit of radius, $r_E \sim 1.5 \times 10^{11}$ [m] (1 [AU]) 
and $r_R \sim 2.28 \times 10^{11}$ [m] (1.52 [AU]), respectively. 
  \begin{figure}
   \begin{center}
   \includegraphics[scale=0.65]{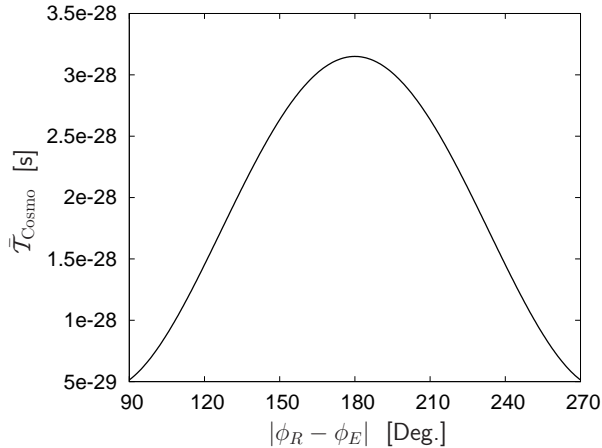}
   \caption{Extra time delay $\bar{\cal T}_{\rm Cosmo}$
    measured in the proper time of observer on the Earth.
    \label{fig-arakida1}}
   \end{center}
  \end{figure}
Figure \ref{fig-arakida1} shows the estimated extra time delay 
$\bar{\cal T}_{\rm Cosmo}$ in the proper time of observer.
The figure is plotted as the function of relative angle between the 
Mars $\phi_{R}$ and the Earth $\phi_E$, 
$|\phi_R - \phi_E|$ for $|\phi_R - \phi_E| > \pi/2$. 
The order of $\bar{\cal T}_{\rm Cosmo}$ is at most about 
$3.2 \times 10^{-28}$ [s] which is considered as the typical 
magnitude of $\bar{\cal T}_{\rm Cosmo}$ for the planetary radar and 
the spacecraft ranging. Here let us evaluated the 
order of magnitude of each part in (\ref{dtau}); 
the leading terms $H^2_0 r^3/c^3 \sim 10^{-28}$, the coupling terms
$GMH^2_0 r^2/c^5 \sim v^2 H^2_0 r^3/c^5 \sim 10^{-36}$, and
$H^4_0r^5/c^5 \sim 10^{-57}$ where we took $r_E, v_E$ of Earth
as $r, v$. However, it is much smaller than the current measurement 
limit, that is, the internal error of atomic clocks, $10^{-9}$ [s].

Next, we calculate (\ref{tauh2}) presuming again the Earth moving 
along the circular orbit, $r_E = 1.5 \times 10^{11} ~[{\rm m}]$.
The difference between the coordinate time and the proper one 
caused by the cosmological expansion is, in the interval of 100 [yr],
\begin{eqnarray}
 \tau_{\rm Cosmo} = 
  - \frac{H_0^2r_E^2}{2c^2}(t - t_0) 
  \sim
  - 4.1 \times 10^{-20} ~~ {\rm [s]}.
  \label{tauh3}
\end{eqnarray}
This is also much smaller than the relative error of atomic clocks, 
$10^{-9}$ [s]. $\tau_{\rm Cosmo}$ causes the apparent orbital 
variations of longitude $\phi$ of planet, but it is 
$\delta \phi = n_0\tau_{\rm Cosmo} \sim -8.2 \times 10^{-27}$ 
[rad] where $n_0 = \sqrt{GM/r_E^3} \sim 2.0\times 10^{-7}$ [rad/s]
is the mean motion of Earth.
Then $\tau_{\rm Cosmo}$ does not has much effect on the observed 
$d{\rm AU}/dt$.
Rounding up the results, the effect of cosmological expansion 
on the light propagation dose not give the explanation of 
observed increase of AU in the current solar system experiments.

Here we mention the attempts of Krasinsky and Brumberg (2004).
They also investigated the secular increase of AU in terms of 
cosmological expansion deriving the approximate metric of Einstein 
equation, (5) of \cite{kb2004} 
which is the modified version of the standard form of 
Robertson-McVittie metric (\ref{mcvittie}).
\footnote{(5) of \cite{kb2004} can be recovered from the
lowest order of (\ref{mcvittie}) with respect to $GM/c^2$, 
replacing the time coordinate $dt \rightarrow a(t)dt$ and 
$a(t) \rightarrow \sqrt{A}$, see appendix A of \cite{kb2004}.}
In such a comoving form, they examined both the planetary motion
and light propagation, then found there exist the large 
orbital variations in the radial and longitudinal directions due to the
cosmological expansion ((19) and (20) of \cite{kb2004}, respectively).
However, these dynamical perturbations are completely canceled out by
the time transformation between the coordinate time and proper one
relating with light/signal propagation ((22) and (23) of \cite{kb2004}, 
respectively). Hence they concluded that the cosmological effect 
cannot account for the observed $d{\rm AU}/dt$.

Before closing this section, we shortly note 
the dynamical effect due to the cosmological terms in the planetary 
motion. This problem has been studied by many authors 
\cite{mcvittie,jarnefelt1,jarnefelt2,jarnefelt3,es1,es2,schucking,
noerd1,gautreau,cooperstock,carrera,ks2004,sereno,faraoni,mashhoon}.
And they showed that the cosmological influence on the planetary
dynamics, e.g. the additional perihelion advance, is negligible 
small. We also confirmed that the cosmological terms are too small to
give an account of the increasing of AU. In appendix 
\ref{append-a}, we briefly summarize the results of the cosmological 
effect on the planetary motion based on (\ref{base-metric}).
\section{Summary}
In this paper, we investigated the light propagation passing near 
the gravitating body as the Sun embedded in the cosmological 
spacetime by means of the Robertson-McVittie solution.
We concentrated on the time delay and derived the correction terms 
with respect to the Shapiro's formula. We also took account of the 
time transformations; coordinate time to proper one, and conversely, 
proper time to coordinate one. These treatments are of importance when 
dealing with the observed round-trip time and its reduction process.

As the application of results, we considered the problem of increase of 
astronomical unit reported by Krasinsky and Brumberg (2004), 
since the AU is currently derived from the analysis of the round-trip 
time of light/signal. However, the effect of the cosmological expansion 
is too small to produce the observed $d{\rm AU}/dt = 15$ [m/century]
and then it dose not give the explanation of increase of AU.

Here it is worth to mention the interpretation of $d{\rm AU}/dt$.
Although we and Krasinsky and Brumberg (2004) investigated the secular 
increasing of astronomical unit in terms of the cosmological effect, 
another possibility is suggested, that is, increasing of AU is 
arisen due to the lack of the calibrations of internal 
delays of the radio signals within the spacecrafts and 
this may be the most plausible reason of observed $d{\rm AU}/dt$.
Nonetheless, till now, the origin of $d{\rm AU}/dt$ is far from 
clear therefore this issue should be explored by means of the 
every possibility we can imagine. And the re-analysis 
of $d{\rm AU}/dt$ using new data sets is also expected.

As we mentioned in the end of section \ref{au}, so far, the dynamical
terms due to cosmological expansion in the planetary dynamics has 
been studied actively, while the influence on light/signal 
propagation has hardly been examined. However, the accuracy of 
astronomical/astrophysical observations, especially the optical, radio
and arrival time ones, rapidly increases in the solar system and they 
may make it necessary to incorporate the cosmological effect in the
future date. For this purpose, our investigation in present paper is 
the first attempt to this direction. To consider this problem,
the matching of BCRS and cosmological reference system discussed by
\cite{ks2004,kopeikin2007} is also important issue to be investigated.

(\ref{base-metric}) expresses the completely static gravitational 
field, and is equivalent to Schwarzschild-de Sitter model.
Nevertheless, as long as we focus on the solar system
experiments, our results may be regarded as the dominant effects 
due to the cosmological expansion, see in (\ref{hubble_0}). 
However in order to study the cosmological influence  
on the light/signal propagation in more detail and general cases, 
we must treat $H(t)$ and/or $a(t)$ as a function of time and adopt 
the time dependent spacetime model. This subject will be 
investigated at some future time.

%

We would like to express our gratitude to the referee for fruitful 
comments and suggestions.
We acknowledge to G. A. Krasinsky for providing the
information and comments about the AU issue. 
We also appreciate T. Fukushima, M. Kasai, H. Asada, Y. Itoh and 
M. Takada for fruitful discussions and comments. 
This work was partially supported by the Ministry of Education, 
Science, Sports and Culture, Grant-in-Aid, 18740165.

\appendix
\section{Planetary Motion Based on Metric (\ref{base-metric})
\label{append-a}}
Cosmological influence on the planetary motion has studied by 
many authors as we mentioned in section \ref{au}. 
Then, in this appendix we limit to summarize our results 
based on (\ref{base-metric}) briefly.

The equation of motion in the quasi-Newtonian form becomes, 
\begin{equation}
 \frac{d^2r}{dt^2} - r\left(\frac{d\phi}{dt}\right)^2 
  + \frac{GM}{r^2} = H_0^2 r,\quad
  \frac{d}{dt}\left(r^2\frac{d\phi}{dt}\right) = 0.
  \label{newton}
\end{equation}
Supposing initially circular orbit $r = r_0 = \mbox{const.}$,
the mean motion is expressed as,
\begin{equation}
  n = \frac{d\phi}{dt} \simeq n_0 + \delta n,
  \quad
  \delta n = - n_0\frac{r_0^3 H_0^2}{2GM} \sim 1.0 \times 10^{-28}
  ~~ {\rm [rad/s]},
\label{mm}
\end{equation}
where $n_0 = \sqrt{GM/r_0^3} \sim 2.0\times 10^{-7}$ [rad/s] and
we took the orbital radius of Earth $r_E$ as $r_0$. 
From (\ref{mm}) the variation of longitude $\phi$ is evaluated 
in 100 [yr] as,
\begin{equation}
 \delta \phi = \delta n (t - t_0) 
  \sim - 3.5 \times 10^{-18} ~~ {\rm [rad]}.
\end{equation} 
Also using (\ref{mm}), the orbital period of planet 
(in this case that of Earth) is written as, 
\begin{eqnarray}
 T = \frac{2\pi}{n} \simeq T_0 + \delta T,\quad
  \delta T = T_0
  \frac{r_0^3 H_0^2}{2GM} \sim 7 \times 10^{-26} ~~{\rm [s]},
\end{eqnarray}
here $T_0 = 2\pi/n_0$. Finally putting $r = r_0 + \delta r$ and 
inserting into (\ref{newton}), it yields,
\begin{equation}
 \delta r = \frac{r_0^4 H_0^2}{GM}
  \left(1 + \frac{r_0^3 H_0^2}{GM}\right) 
  \sim 2.0 \times 10^{-11} ~~ [{\rm m}].
  \label{delta_r}
\end{equation} 
Therefore we find that the estimated orbital variations 
$\delta r, \delta \phi$ and change of orbital period $\delta T$
are much smaller than the observed $d{\rm AU}/dt = 15$ [m/century].

\end{document}